____________________________________________________________________________
\documentstyle[12pt]{article}
\title{Levinson's Theorem for the Klein-Gordon Equation in Two Dimensions}

\author{Shi-Hai Dong\thanks{Electronic address: DONGSH@BEPC4.IHEP.AC.CN}\\
{\scriptsize Institute of High Energy Physics, P. O. Box 918(4), 
Beijing 100039, The People's Republic of China}\\
\\
Xi-Wen Hou\\
{\scriptsize Institute of High Energy Physics, 
P. O. Box 918(4), Beijing 100039}\\
{\scriptsize and Department of Physics, University of Three
Gorges, Yichang 443000, The People's Republic of China}\\
\\
Zhong-Qi Ma\\
{\scriptsize China Center for Advanced Science and Technology
(World Laboratory), P. O. Box 8730, Beijing 100080}\\
{\scriptsize  and Institute of High Energy Physics, P. O. Box 918(4) 
Beijing 100039, The People's Republic of China}}

\date{}

\begin{document}

\maketitle

\begin{abstract}
The two-dimensional Levinson theorem for the Klein-Gordon equation
with a cylindrically
symmetric potential $V(r)$ is established. It is shown that
$N_{m}\pi=\pi \left(n_{m}^{+}-n_{m}^{-}\right)=
[\delta_{m}(M)+\beta_{1}]-[\delta_{m}(-M)+\beta_{2}]$
,where $N_{m}$ denotes the difference between the number of bound
states of the particle $n_{m}^{+}$ and the ones of 
antiparticle $n_{m}^{-}$ with a fixed angular momentum $m$, and
the $\delta_{m}$ is named phase shifts. The 
constants $\beta_{1}$ and $\beta_{2}$
are introduced to symbol the critical cases
where the half bound states occur at $E=\pm M$.
\end{abstract}

\vskip 0.5cm


\begin{center}
\section*{I. Introduction}
\end{center}

The Levinson theorem[3], an important 
theorem in scattering theory, 
established the relation between the 
total number of bound states and the
phase shifts at zero momentum. During 
the past half century, 
the Levinson theorem has been proved 
by several authors with different methods, and generalized 
to different fields [4-11]. Rough speaking, 
there are three main methods  
used to prove the Levinson theorem. One [3] is 
based on the elaborate analysis 
of the Jost function first introduced by Jost. 
The second is relied on the Green function method [7].
The third method is used to demonstrate the 
Levinson theorem 
by the Sturm-Liouville 
theorem [8-10]. This simple, intuitive 
method is readily to be 
generalized and has been verified by 
the proofs of many physical problems
[8-10,22-24]. Furthmore, some obstacles and 
ambiguities, which may occur 
in other two methods, disappear in the third method. 
However, it is found in the later proof 
that the Sturm-Liouville 
theorem can't be directly used to prove 
the Levinson theorem for the
Klein-Gordon equation, but a modified 
method which is similar to the
Sturm-Liouville theorem will be applied 
to prove the Levinson theorem.
Consequently, such a generalization may 
be useful for the method of 
bosonization method which has been widely utilized in the
literature[26].

The Klein-Gordon equation, which 
describes the motion of a relativistic
scalar particle, is a second-order differential 
equation with respect to both space
and time. When there exists a potential as the fourth
component of the vector field, the energy 
eigenvalues are not necessarily real
and the eigenfunctions satisfy the orthogonal 
relations with a weight
factor[1-2] such that a parameter ${\bf \epsilon}$ which 
is not always real and positive
appears in the normalized relation with a weight factor. 
As pointed out by Pauli and Snyder ${\it at~al}$[1-2]
,after bose quantization, that 
those amplitudes with real and 
positive ${\bf \epsilon} $ describe 
particles, but those with real and 
negative ${\bf \epsilon} $ antiparticles.

Recalling in the three-dimensional spaces, two 
main methods are used to set up the Levinson theorem 
for the Klein-Gordon equation. One is relied on some
formulae which are valid for the cases without 
complex energies[7]. The other, which is similar to 
that of Sturm-Liouville theorem, is applied to
arrive at the Levinson theorem for the Klein-Gordon equation[9]. 
This result is correct for the cases both without complex
energies and  with complex energies.

The reasons why we write this 
paper are that, on the one hand
the Levinson theorem in two dimensions has been studied 
in experiment [19] as well as in theory [20-24] in virtue of 
the wide interest in lower-dimensional field theories and 
other modern physics [12-18], on the other hand the Levinson 
theorem for the Klein-Gordon equation in two
dimensions has never been appeared in the literature. 
In our previous works[22-24],
some surprised results are obtained 
from the nonrelativistic and relativistic
particle as well as the non-local interactions 
in two dimensions. We attempt
to set up the Levinson theorem for the 
Klein-Gordon equation in two
dimensions. What new results will be appeared?

This paper is organized as follows. In Sec. II, we 
review the properties of
the Klein-Gordon equation, especially those related with the 
parameter ${\bf \epsilon} $. 
In Sec. III, it is proved that the difference 
between the numbers of bound states of particle and the ones
of antiparticle   
only relies on the changes of the 
logarithmic derivatives of the wave 
functions at $E=\pm M$ as the 
potential $V(r)$ changes from zero to the 
given value. In Sec. IV, it is also turned 
out that these changes 
are closely connected with the 
phase shifts at $E=\pm M$ which then results in the 
establishment of the two-dimensional Levinson 
theorem for the Klein-Gordon equation.

\begin{centering} 
\section*{II. the Klein-Gordon Equation}
\end{centering}

Throughout this paper the natural units $\hbar=c=1$ are employed.
Consider a relativistic scalar 
particle satisfying the Klein-Gordon equation
$$\left(-\nabla ^{2}+M^2\right)\psi(x)
=\left[E-V(x)\right]^{2}\psi(x),\eqno(1)$$
where the potential $V(x)$ is 
the fourth component of a vector field and the
$M,E$ denote the mass and the 
energy of the particle, respectively. In order
to simplify the discussion, we 
only research that the potential is static
and cylindrical symmetric one
$$V(x)=V(r),\eqno(2)$$
and its asymptotic behavior is written
$$r|V(r)|\rightarrow 0~~~~
{\rm when}~~~~ r\longrightarrow 0,\eqno(3a)$$

\noindent
and
$$V(r)=0~~~{\rm when}~~r>r_{0}.\eqno(3b)$$
Equation (3a) is required to make the wave 
function single value at the origin, and (3b) is called  
the cutoff potential for the sake of the simplicity of discussion,
i.e it is vanishing beyond a 
sufficiently large radius $r_{0}$. It is
proved that, following the 
method [22-23], the results obtained in this paper
will not change the essence of the 
proof if the potential vanishes faster
than $r^{-2}$ at infinity. 

Introduce a parameter $\lambda$ for the potential $V(r)$
$$V(r,\lambda)=\lambda V(r), \eqno (4) $$

\noindent
which shows that the potential
$V(r,\lambda)$ changes from zero to the given potential
$V(r)$ when $\lambda$ increases from zero to one

\noindent
Due to the symmetry of the potential, Let
$$\psi(\bf x,\lambda)=r^{-1/2} R_{m}(r,\lambda) e^{ \pm im\varphi},
~~~~~m=0, 1, 2, \ldots,  \eqno (5) $$  

\noindent
where the radial wave equation 
$R_{m}(r,\lambda)$ satisfies the radial
equation
$$\displaystyle {\partial^{2}  R_{m}(r,\lambda) \over \partial r^{2} }
+\left\{(E^2-M^2)-(2EV-V^2)-
\displaystyle{\frac{m^2-1/4}{r^2}}\right\}
R_{m}(r,\lambda)=0.\eqno(6)$$

\noindent
Denote by ${R}_{m1}(r,\lambda)$ the solution to Eq.(6) for the 
energy $E_{1}$
$$\displaystyle {\partial^{2}R_{m1}(r,\lambda) \over \partial
r^{2}}
+\left\{(E_{1}^2-M^2)-(2E_{1}V-V^2)-
\displaystyle {m^{2}-1/4 \over r^{2}} \right\} 
R_{m1}(r,\lambda)=0. \eqno (7) $$

\noindent
Multiplying Eq.(6) and Eq.(7) by $R_{m1}(r,\lambda)$ and
$R_{m}(r,\lambda)$, respectively, and calculating their difference,
we have
$$\displaystyle {\partial \over \partial r} \left\{ R_{m}(r,\lambda)
R_{m1}^{\prime \ast}(r,\lambda) 
-R_{m1}(r,\lambda) R_{m}^{\prime \ast}(r,\lambda)\right\}
=-(E_{1}^{\ast}-E)R_{m1}^{\ast}(r,\lambda)~\cdot~(E_{1}^{\ast}+E-2V)
R_{m}(r,\lambda),\eqno (8) $$

\noindent
where the primes denote the derivative 
of the radial wave function with
respect to the variable $r$. 
As we know, the energy eigenvalues are not
necessarily real for some potential $V(r)$ which 
origins from the Klein
paradox. Integrating (8) over the whole space and noting that
$R_{m}(r,\lambda)R_{m1}^{\prime \ast}(r,\lambda)-
R_{m1}(r,\lambda)R_{m}^{\prime \ast}(r,\lambda)$
vanishes both at the origin and at 
infinity for the physically admissible
solutions with the different 
energies $E$ and $E_{1}$, we get the weighted
orthogonality relation of the radial wave function
$$(E_{1}^{\ast}-E)\int_{0}^{\infty}
R_{m1}^{\ast}(r,\lambda)(E_{1}^{\ast}+E-2V)R_{m}(r,\lambda)dr=0.\eqno(9)$$

\noindent
As a matter of fact, we always able to obtain the ${\it real}$
solutions for the ${\it real}$ energies.
However, it is easy to see from Eq. (9) that the 
normalized relation for the
solutions with real energies are not 
always positive on account of the
weight factor $(E_{1}^{\ast}+E-2V)$:
$$\int_{0}^{\infty}R_{m1}(r,\lambda)(E_{1}+E-2V)R_{m}(r,\lambda)dr
=\left\{\begin{array}{ll}
{\bf \epsilon_{E}}\delta(E_{1}-E),&|E|>M,\\
{\bf \epsilon_{E}}\delta_{E_{1}E},&|E|<M.
\end{array}
\right.\eqno(10)$$

\noindent
The parameter ${\bf \epsilon}_{E}$, which 
depends on the particular radial wave function $R_{m}(r,\lambda)$,
may be either positive, negative or vanishing. 
Normalized factors of the
solutions can't change the sign of ${\bf \epsilon}$. 
Generally speaking, if the
solution $R_{m}(r,\lambda)$ with a complex 
energy $E$ is complex, then $R_{m}^{\ast}$
is also a solution with complex energy $E^{\ast}$ and 
a complex ${\bf \epsilon}_{E}$
appears for a pair of the complex solutions. It 
is evident after bose
quantization that those $R_{m}(r,\lambda)$ with 
positive ${\bf \epsilon}_{E}$ describes
particles and those with negative ${\bf \epsilon}_{E}$ 
antiparticles. In the case
zero ${\bf \epsilon}_{E}$, the solution can 
be regarded as a pair of particle and
antiparticle bound states. The Hamiltonian 
and charge operator can't be
written as the diagonal forms for the 
solutions with complex energy
${\bf \epsilon}_{E}$, therefore they 
describe neither particles nor antiparticles.
In this paper, we only count the number 
of bound states with the real positive
and negative nonvanishing ${\bf \epsilon}_{E}$ 
is named particle and antiparticle
bound states, respectively.

Since we are always able to arrive at the
${\it real}$ solution for the ${\it real}$ energy,
we can now solve Eq.(6) in two regions and
match two solutions at $r_{0}$. Actually, 
the solutions in the region
$[0,r_{0}]$ with $R_{m}(0)=0$ can be arrived at in principle. 
We only need one matching condition at $r_{0}$ for 
the logarithmic derivative of the radial wave function
$$A_{m}(E,\lambda)\equiv \left\{ \displaystyle {1 \over R_{m}(r,\lambda) }
\displaystyle {\partial R_{m}(r,\lambda) \over \partial r}\right\}_{r=r_{0}-} 
=\left\{ \displaystyle {1 \over R_{m}(r,\lambda) }
\displaystyle {\partial R_{m}(r,\lambda) \over \partial r}
\right\}_{r=r_{0}+} \equiv B_{m}(E). 
\eqno (11) $$

Only one solution is convergent at the
origin because of the condition (3a). For example, for the free 
particle ($\lambda=0$), the solution
to Eq. (6) at the region $[0,r_{0}]$ is proportional to the 
Bessel function $J_{m}(x)$:
$$R_{m}(r,0)=\left\{\begin{array}{ll}
\sqrt{\displaystyle {\pi kr \over 2 }}J_{m}(kr),~~~~~
&{\rm when}~~|E|>M~~{\rm and}~~k=\sqrt{E^2-M^2} \\
e^{-im\pi/2}\sqrt{\displaystyle {\pi \kappa r \over 2 }}J_{m}(i\kappa r)
,~~~~~&{\rm when}~~|E|<M~~{\rm and}~~\kappa=\sqrt{M^2-E^2}, 
\end{array} \right. \eqno (12) $$

\noindent
The solution $R_{m}(r,0)$ given in Eq. (12) is a real function. 
A constant factor on the radial wave function $R_{m}(r,0)$ is not 
important.

In the region $[r_{0},\infty)$, we have $V(r)=0$. For $|E|>M$, 
there are two oscillatory solutions to Eq. (6). Their combination 
can always satisfy the matching condition (11), so that there
is a continuous spectrum for $|E|>M$.
$$R_{m}(r,\lambda)=\sqrt{\displaystyle {\pi kr \over 2 }}
\left\{ \cos \eta_{m}(k,\lambda)J_{m}(kr)-\sin \eta_{m}(k,\lambda)
N_{m}(kr) \right\}~~~~~~~~~~~~~~~~~$$
$$~~~~~~~~ \sim \cos \left(kr-\displaystyle{m\pi \over 2}-
\displaystyle {\pi \over 4} +\eta_{m}(k,\lambda) \right),~~~~~~~~~~~~~
{\rm when}~~r\longrightarrow \infty.  \eqno (13) $$

\noindent
where $N_{m}(kr)$ is the Neumann function.

However, there is only one convergent solution 
in the region $[r_{0},\infty)$
for $|E|\leq M$ the matching condition (11) is not always satisfied. 
$$R_{m}(r,\lambda)=e^{i(m+1)\pi/2}\sqrt{\displaystyle 
{\pi \kappa r \over 2 }}
H^{(1)}_{m}(i\kappa r) \sim e^{-\kappa r}, ~~~~~
{\rm when}~~r\longrightarrow \infty.  \eqno (14) $$

\noindent
where $H^{(1)}_{m}(x)$ is the Hankel function of the first kind.
When the condition (11) is satisfied, a bound state appears
at this energy. It means that there is a discrete spectrum 
for $|E|\leq M $.

\noindent
As mentioned above, in the case with the ${\it real}$ energy solutions, 
integrating the Eq. (6) in two regions $[0,r_{0}]$ and $[r_{0},\infty)$
,respectively, and taking the limit $E_{1}\rightarrow E$, we will obtain the
following equations in terms of the boundary condition that $R_{m}(0)=0$ 
and $R_{m}(\infty)=0$ for $|E|<M$ 
$$\begin{array}{rl}
\displaystyle {\partial A_{m}(E,\lambda) \over \partial E}&\equiv~ 
\displaystyle {\partial \over \partial E} \left(
\displaystyle {1 \over R_{m}(r,\lambda)}
\displaystyle {\partial R_{m}(r,\lambda)\over \partial r}
\right)_{r=r_{0}-} \\
&=~-R_{m}(r_{0},\lambda)^{-2}
\displaystyle \int_{0}^{r_{0}}R_{m}(r,\lambda)^{2}~2~[E-V(r)]~dr<0.
\end{array} \eqno (15a) $$

\noindent
and

$$\begin{array}{rl}
\displaystyle {dB_{m}(E) \over dE}&\equiv~
\displaystyle {\partial \over \partial E} \left(
\displaystyle {1 \over R_{m}(r,\lambda)}
\displaystyle {\partial R_{m}(r,\lambda)\over \partial r}
\right)_{r=r_{0}+}\\
&=~R_{m}(r_{0},\lambda)^{-2}
\displaystyle \int_{r_{0}}^{\infty}R_{m}(r,\lambda)^{2}2~E~dr~>0.
\end{array} \eqno (15b) $$

\noindent
which demonstrates from Eq. (15) that $A_{m}(E,\lambda)$ is no longer
monotonic with respect to energy, but $B_{m}(E)$ is still monotonic with
respect to energy if the energy doesn't change sign.

From the matching condition (11) we have
$$\tan \eta_{m}(k,\lambda)=\displaystyle {J_{m}(kr_{0}) \over
N_{m}(kr_{0})} ~\cdot ~\displaystyle
{A_{m}(E,\lambda)-kJ'_{m}(kr_{0})/
J_{m}(kr_{0})-1/(2r_{0})
\over A_{m}(E,\lambda)-kN'_{m}(kr_{0})/
N_{m}(kr_{0})-1/(2r_{0}) }.  \eqno (16) $$
$$\eta_{m}(k)\equiv \eta_{m}(k,1). \eqno (17) $$ 

\noindent
where the prime denotes the derivative of the Bessel function,
the Neumann function, and later the Hankel function with respect
to their argument. However, it is not true for $|E|<M$ because of no
adjustable phase shift $\delta_{m}(E)$. Once the matching condition is
satisfied, we will get the discrete bound states.

The phase shift $\eta_{m}(k,\lambda)$ is determined from (16)
up to a multiple of $\pi$ due to the period of the tangent function.
In this paper, for the free particle $(V(r)=0)$, 
the definition of phase shift 
$\eta_{m}(k,0)$ is defined to be zero, i.e
$$\eta_{m}(k,0)=0,~~~~{\rm where}~~\lambda=0, \eqno (18) $$
which is same as our previous definition[8-9,22-24].

It is shown from Eq. (10) that scattering states----$|E|>M$----are 
normalized as the Dirac $\delta$ function, and that 
the main contribution to the integration Eq. (10) 
comes from the radial wave functions
in the region$[r_{0},\infty)$ where there 
is no potential. For this reason we obtain
$$\epsilon_{E}=\pi \sqrt{E^2-M^2}~\cdot~\frac {E}{|E|},~~~|E|>M.\eqno(19)$$

\noindent
All the scattering states with positive 
energy $(E>M)$ describe particles
and those with negative energy $(E<-M)$ 
describe antiparticles. It is
easy to see that this conclusion is not 
true for the critical case $E=\pm M$
except for $S$ waves where there is a half 
bound state at $E=\pm M$.
The situations which ${\bf \epsilon}_{E}$ 
with $E=\pm M$ and $m>1$ 
may be positive, negative or vanishing are relied on
the potential.

\begin{centering}
\section*{III. The Number of Bound States}
\end{centering}

In our previous works, the Levinson theorem 
for the nonrelativistic and
relativistic particles are set up under 
the help of Sturm-Liouville theorem. 
For the Sturm-Liouville
problem, the fundamental trick is the definition of a phase
angle which is monotonic with 
respect to the energy [25]. Although this
method is very simple, intuitive 
and easy to be generalized, from
the Eq. (6), it is the weight 
factor $(E_{1}^{\ast}+E-2V)$ that makes the
Sturm-Liouville theorem not be 
used for the the Klein-Gordon equation.
Nevertheless, a modified method is applied to 
prove the Levinson theorem 
for the Klein-Gordon equation.
From the difference between the 
Eq.(15a) and Eq. (15b), we arrive at
$$\displaystyle{\frac{dB_{m}(E)}{dE}}-
\displaystyle{\frac{\partial A_{m}(E,\lambda)}{\partial E}} 
\equiv B'_{m}(E)-A'_{m}(E,\lambda)=
\displaystyle{\frac{1}{R_{m}(r_{0})}}{\bf \epsilon}_{E},
\eqno(20)$$

\noindent
where here and hereafter the primes 
denote the derivative with respect to
the energy.

From Eq. (14), we get
$$B_{m}(E)=\displaystyle {i\kappa H^{(1)}_{m}(i\kappa r_{0})' 
\over H^{(1)}_{m}(i\kappa r_{0}) }-\displaystyle {1 \over 2r_{0}}
=\left\{\begin{array}{ll} (-m+1/2)/r_{0}\equiv \rho_{m}
&{\rm when}~~k_{1}\longrightarrow 0 \\
-\kappa \sim -\infty &{\rm when}~~k_{1}\longrightarrow \infty.
\end{array} \right. \eqno (21) $$

\noindent
The logarithmic derivative given in Eq. (21) does not depend
on $\lambda$. On the other hand, when $\lambda=0$ we obtain 
from Eq. (12)
$$A_{m}(E,0)=\displaystyle {i\kappa J'_{m}(i\kappa r_{0}) 
\over J_{m}(i\kappa r_{0}) }-\displaystyle {1 \over 2r_{0}}
=\left\{\begin{array}{ll} (m+1/2)/r_{0} 
&{\rm when}~~k_{1}\longrightarrow 0\\
\kappa \sim \infty &{\rm when}~~k_{1}\longrightarrow \infty.
\end{array} \right. \eqno (22) $$

\noindent
It is evident from the Eqs. (21) and (22) that 
both $B_{m}(E)$ and $A_{m}(E,0)$ are
continuous curves with respect to energy 
which don't intersect each other;
i.e. the matching condition (11) is 
not satisfied if $|E|\leq M$ 
and $\lambda =0$. No bound states 
appear when there is no potential.

As $\lambda$ changes from the zero to the 
given potential, $B_{m}(E)$ don't
change, but $A_{m}(E,\lambda)$ changes 
continuously except the points where
$R_{m}(r_{0})=0$ and $A_{m}(E,\lambda)$ tends 
to infinity. Generally speaking,
$A_{m}(E,\lambda)$ is continuous except 
those finite points and intersects
with the curve $B_{m}(E)$ several times 
for $|E|\leq M$. The bound state will
appear only if the intersection happens. 
The points of the intersection
determine the number of the bound states. It 
is shown from Eq. (20) that 
the relative slopes at the points of 
intersection decide whether the 
bound states describe particle or antiparticles.

When the potential $V(r)$ change 
with the $\lambda$, the number of
intersection points will change, too. This 
only origins from the following two
sources. Firstly, the intersection points 
move inward or outward at $E=\pm M$. Secondly, the
curve $A_{m}(E,\lambda)$ intersects with 
the curve $B_{m}(E)$ or departs from it
through the tangency point. For the second 
case, a pair of particle
and antiparticle bound state will be 
created or annihilated at the same time,
but the difference of the number of the particle $n_{m}^{+}$
and antiparticle bound state $n_{m}^{-}$ don't 
change. That's to say, the change of 
the whole bound states $N_{m}$ 
which expresses that the difference of 
the particle bound state and the
antiparticle state only depends on 
the intersection points moving 
in or out at $E=\pm M$ where the critical cases
occur. Hence, we only discuss this case. 
There are four cases when
$A_{m}(M,\lambda)=B_{m}(M)$ when $A_{m}(M,\lambda)$ 
decreases across the value
$B_{m}(M)=(-m+1/2)/r_{0}$ at $E=M$
$$(1)A'_{m}(M,\lambda)<B'_{m}(M),\eqno(23a)$$
$$(2)A^{(n)}_{m}(M,\lambda)=B^{(n)}_{m}(M),~~
(-1)^{n}A^{(n+1)}(M,\lambda)<(-1)^{n}B^{(n+1)}_{m}(M),\eqno(23b)$$
$$(3)A^{(n)}_{m}(M,\lambda)=B^{(n)}_{m}(M),~~
(-1)^{n}A^{(n+1)}_{m}(M,\lambda)>(-1)^{n}B^{(n+1)}_{m}(M),\eqno(23c)$$
$$(4)A'_{m}(M,\lambda)>B'(M).\eqno(23d)$$

Where here and hereafter $n$ is positive integer.
For the first two situation, a interaction point 
moves inward from $E>M$ to $E<M$,
which results in the appearance of new particle 
bound state. However, for
the last two cases, the interaction point 
moves outward from $E<M$ to $E>M$,
which causes the disappearance of an 
antiparticle bound state. The converse
process occurs when $A_{m}(M,\lambda)$ 
increases to cross the value $B_{m}(M)$,
i.e. the number of bound 
states $N_{m}$ increases by one only if
each time $A_{m}(M,\lambda)$ decreases to 
cross the value $B_{m}(M)$ at $E=M$.
Conversely, each time $A_{m}(M,\lambda)$ increases 
across the value $B_{m}(M)$ 
at $E=M$, $N_{m}$ decreases by one.

On the other hand, there are also four cases 
when $A_{m}(-M,\lambda)=B_{m}(-M)$:
$$(1')A'_{m}(-M,\lambda)>B'_{m}(-M),\eqno(24a)$$
$$(2')A^{(n)}_{m}(-M,\lambda)=B^{(n)}_{m}(-M),
A^{(n+1)}_{m}(-M,\lambda)>B^{(n+1)}_{m}(-M),\eqno(24b)$$
$$(3')A^{(n)}_{m}(-M,\lambda)=B^{(n)}_{m}(-M),
A^{(n+1)}(M,\lambda)<B^{(n+1)}(M),\eqno(24c)$$
$$(4')A'_{m}(-M,\lambda)<B'_{m}(-M).\eqno(24d)$$

\noindent
If $A_{m}(-M,\lambda)$ decreases across the 
value $B_{m}(-M)$ as $\lambda$
increases, for the first two cases a interaction moves inward
from the $E<-M$ to $E>-M$ point which describes an
antiparticle. But for the last two cases 
an interaction point moves outward
from $E>-M$ to $E<-M$ which describes a particle.
The number of bound states $N_{m}$ 
decreases by one only if each time
$A_{m}(-M,\lambda)$ increases across 
the value $B(-M)$. The opposite process
occurs when $A_{m}(-M,\lambda)$ increases 
across the value $B_{m}(-M)$.

We denote by $N_{m}(\pm M)$ the difference 
between the number of 
times $A(\pm M,\lambda)$ decreasing
across the value $B(\pm M)$ and the 
number of the times that
$A(\pm M,\lambda)$ increasing across that 
value. Hence, we obtain
$$N_{m}\equiv n_{m}^{+}-n_{m}^{-}=n_{m}(+M)-n_{m}(-M).\eqno(25)$$

\begin{centering}
\section*{IV. The Phase Shifts}
\end{centering}

As we know, the solutions in the region $[r_{0},\infty)$ for the scattering
states have been given by Eq. (13). 
The phase shift $\eta_{m}(0,\lambda)$ is the limit of the phase shift 
$\eta_{m}(k,\lambda)$ as $k$ tends to zero. Hence, what we are 
interested in is the phase shift $\eta_{m}(k,\lambda)$ at a
sufficiently small momentum $k$, $k\ll 1/r_{0}$. For the small momentum
we obtain from the matching condition (11)

$$\begin{array}{l}
\tan \eta_{m}(k,\lambda)\sim \\[2mm]
\sim\left\{\begin{array}{ll}
\displaystyle  {-\pi (kr_{0})^{2m} \over 2^{2m}m!(m-1)!}
~\cdot~\displaystyle  {A_{m}(0,\lambda)-(m+1/2)/r_{0} \over 
A_{m}(0,\lambda)-c^{2}k^{2}-\rho_{m} \left(1-
\displaystyle  {(kr_{0})^{2} \over (m-1)(2m-1) }\right) } 
&{\rm when}~~m \geq 2 \\
\displaystyle  {-\pi (kr_{0})^{2} \over 4}~\cdot~
\displaystyle  {A_{m}(0,\lambda)-3/(2r_{0}) \over 
A_{m}(0,\lambda)-c^{2}k^{2}-\rho_{1}
\left(1+ 2(kr_{0})^{2}\log(kr_{0})\right) } 
&{\rm when}~~m = 1 \\
\displaystyle  {\pi \over 2\log(kr_{0})}~\cdot~
\displaystyle  {A_{m}(0,\lambda)-c^{2}k^{2}-\rho_{0} \left(
1-(kr_{0})^{2} \right) \over
A_{m}(0,\lambda)-c^{2}k^{2}-\rho_{0}\left(1+
\displaystyle  {2 \over \log (kr_{0})} \right) }
 &{\rm when}~~m =0. \end{array}\right. \end{array} \eqno (26) $$

\noindent
In addition to the leading terms, we include in (26) 
some next leading terms, which is useful only for the 
critical case where the leading terms are canceled with each other.

\noindent
and

$$\left. \displaystyle {\partial \eta_{m}(k,\lambda) \over 
\partial A_{m}(E,\lambda)}\right|_{k}
=\displaystyle {-8r_{0}\cos^{2}\eta_{m}(k,\lambda) \over
\pi \left\{2r_{0}A_{m}(E,\lambda)N_{m}(kr_{0})-2kr_{0}N'_{m}(kr_{0})-
N_{m}(kr_{0})\right\}^{2} } \leq 0, \eqno (27) $$

\noindent
which shows that the phase shift is monotonic with respect to the
logarithmic derivative $A_{m}(E,\lambda)$ as $\lambda$ increases.

It is shown from Eqs. (26) and (27) that they are 
not different from those of
Schr\"{o}dinger equation. Therefore, we may simply discuss 
this problem by the same
method. Each time $A_{m}(\pm M,\lambda)$ decreases across the 
value $B_{m}(\pm M)$ as
the potential changes from the zero to the given potential, 
the phase shift
$\delta_{m}(\pm M,\lambda)$ increases by $\pi$. 
Conversely, the phase shift
$\delta_{m}(\pm M,\lambda)$ decreases by $\pi$ 
if $A_{m}(\pm M,\lambda)$ increases across the value $B_{m}(\pm M)$.

As $\lambda$ increases from zero to one, i.e. the potential changes from the
zero to the given value, we have 
$$\delta_{m}(\pm M)\equiv \delta_{m}(\pm M,1)=n_{m}(\pm M)\pi,\eqno(28)$$
Thus, we draw a conclusion that the Levinson theorem for the Klein-Gordon
equation if $A(\pm M,1)\not =B(\pm M)$
$$N_{m}\pi=\delta_{m}(M)-\delta_{m}(-M).\eqno(29)$$

We now discuss the critical cases
$$A_{m}(M,1)=B_{m}(M)~~~~{\rm and}~~~~A_{m}(-M,1)=B_{m}(-M),\eqno(30)$$

\noindent
where the potential changes from the zero 
to the given potential $V(r)$.
Similar to the discussion[22-24], the phase 
shift $\delta_{m}(\pm M,\lambda)$
increases by $\pi$ for $m>1$ or an additional $\pi$ for the $P$ waves if
$A_{m}(\pm M,\lambda)$ decreases from near 
and larger than the value $B_{m}(M)$ to
smaller than that value when the potential 
changes from the zero to the
given potential. Conversely, 
$\delta_{m}(\pm M,\lambda)$ doesn't decrease by
$\pi$ or an additional $\pi$ if $A_{m}(\pm M,\lambda)$ 
increases across the
value $B_{m}(M)$ as the potential changes to 
the given potential $V(r)$. On the
other hand, the states for $m=0,1$ are called 
a half bound state which is
defined as its wave function is finite but 
not square integrable. Furthermore,
the half bound state is not a bound state. 
For $M>1$ states in the critical
situations, there is a bound state but 
its ${\bf \epsilon}_{E}$ may be either
positive, negative or vanishing, which 
depends on the different cases (23)
and (24). We consider the state with 
zero ${\bf \epsilon}_{E}$ as a pair of particle
and antiparticle bound states.

Introduce two parameters $\beta_{1},\beta_{2}$ 
to describe the appearance or
disappearance of the bound states at the 
critical cases. $\beta_{1}=0$ for
the noncritical case $A(M,1)\not=B_{m}(M)$, 
and $\beta_{2}=0$ for the case
$A_{m}(-M,1)\not=(-M)$.

\noindent
(1) If $A(M,1)=B_{m}(M),\beta_{1}=0$ for 
the cases (23a) or (23c) with
$m>1$;$\beta_{1}=-1$ for the cases (23b) or (23d) with $m>1$; and
$\beta_{1}=-1$ for the case (23a) with $m=1$.\\
(2) If $A_{m}(-M,1)=B_{m}(-M),\beta_{1}=0$ 
for the cases (24a) or (24c) with
$m>1$;$\beta_{1}=-1$ for the cases (24b) or (24d) with $m>1$; and
$\beta_{1}=-1$ for the case (24a) with $m=1$, 
where $\lambda$ is substituted
by one. Then, the Levinson theorem for 
the Klein-Gordon equation with the
cylindrical symmetric potential $V(r)$ 
satisfying the asymptotic behavior(3)
$$N_{m}\pi=\pi \left(n_{m}^{+}-N_{m}^{-}\right)
=[\delta_{m}(M)+\beta_{1}]-[\delta_{m}(-M)+\beta_{2}].\eqno(31)$$

According to the above discussion, it is 
easy to find, compared with the case
in the three-dimensional spaces, that the 
phase shifts for the critical sates
changes by an additional $\pi$ not 
by $\pi/2$. This conclusion is same
as the relativistic and 
nonrelativistic particles.

\vspace{10mm}
{\Large \bf  Acknowledgments}. This work was supported by the National
Natural Science Foundation of China and Grant No. LWTZ-1298 of
the Chinese Academy of Sciences.


\begin{thebibliography}{99}

\bibitem{1} W. Pauli, princeton mimeographed notes(1935).
\bibitem{2} H. Snyder and J. Weinberg, Phys. Rev. {\bf 15} 307 (1940); L.I.
Schiff, H. Snyder and J. Weinberg,{\it ibid} {\bf 15} 315 (1940).
\bibitem{3} N. Levinson, K. Danske Vidensk. Selsk. Mat-fys. Medd. {\bf 25}, 
No. 9 (1949). 
\bibitem{4} R. G. Newton, J. Math. Phys. {\bf 1}, 319 (1960);
{\it ibid} {\bf 18}, 1348, 1582 (1977);
{\it Scattering theory of waves and particles}, (Springer-Verlag, 
New York, 2nd ed.,  1982) and references therein. 
\bibitem{5} J. M. Jauch. Helv. Phys. Acta {\bf 30}, 143 (1957). 
\bibitem{6} A. Martin, Nuovo Cimento {\bf 7}, 607 (1958). 
\bibitem{7} G. J. Ni, Phys. Energ. Fort. Phys. Nucl. {\bf 3}, 
432 (1979); Z. Q. Ma and G. J. Ni, Phys. Rev. {\bf D31}, 1482 (1985). 
\bibitem{8} Z. Q. Ma, J. Math. Phys. {\bf 26}(8), 1995 (1985). 
\bibitem{9} Z. Q. Ma, Phys. Rev. {\bf D32}, 2203 and 2213 (1985). 
\bibitem{10} Z. R. Iwinski, L. Rosenberg, and L. Spruch, 
Phys. Rev. {\bf 31}, 1229 (1985). 
\bibitem{11} N. Poliatzky, Phys. Rev. Lett. {\bf70}, 2507
(1993); R. G. Newton, Helv. Phys. Acta {\bf 67}, 20 (1994);
Z. Q. Ma, Phys. Rev. Lett. {\bf 76}, 3654 (1996). 
\bibitem{12} Z. R. Iwinski, L. Rosenberg, and L. Spruch, 
Phys. Rev. {\bf A 33}, 946 (1986); L. Rosenberg, and L. Spruch, 
Phys. Rev. {\bf A 54}, 4985 (1996). 
\bibitem{13} R. Blankenbecler and D. Boyanovsky, Physica {\bf 18D}, 367(1986). 
\bibitem{14} A. J. Niemi and G. W. Semenoff, Phys. Rev. D{\bf 32}, 471(1985). 
\bibitem{15} F. Vidal and J. Letourneaux, Phys. Rev. {\bf C 45}, 418(1992).
\bibitem{16} K. A. Kiers, W. van Dijk, J. Math. Phys. {\bf 37}, 6033 (1996).
\bibitem{17} M. S. Debianchi, J. Math. Phys., {\bf 35}, 2719 (1994).
\bibitem{18} P. A. Martin and M. S. Debianchi, Europhys. Lett. {\bf
34}, 639 (1996).
\bibitem{19} M. E. Portnoi and I. Galbraith, Solid State Commun.
{\bf 103}, 325 (1997).
\bibitem{20} D. Boll\'{e}, F. Gesztesy, C. Danneels, and S. F. 
J. Wilk, Phys. Rev. Lett. {\bf 56}, 900 (1986).
\bibitem{21} Q. G. Lin, Phys. Rev. {\bf A56}, 1938 (1997). 
\bibitem{22} Shi-Hai Dong, Xi-Wen Hou and Zhong-Qi Ma, Levinson's 
theorem for the Schr\"{o}dinger equation in two dimensions, 
Phys. Rev. A. accepted (will be published in August of 1998).
\bibitem{23} Shi-Hai Dong, Xi-Wen Hou and Zhong-Qi Ma, The relativistic
Levinson's theorem in two dimensions, preprint, submitted to Phys. Rev. A.
\bibitem{24} Shi-Hai Dong, Xi-Wen Hou and Zhong-Qi Ma, Levinson's theorem
for the non-local interactions in two dimensions, accepted to J. Phys. A. 
\bibitem{25} C. N. Yang, in {\it Monopoles in Quantum Field Theory}, 
Proceedings of the Monopole Meeting, Trieste, Italy, 1981, 
ed. by N. S. Craigie, P. Goddard, and W. Nahm (World Scientific, 
Singapore, 1982), p.237.
\bibitem{26} C. G. Callan, Jr., Phys. Rev. {\bf 26}, 2058 (1982); E. Witten,
Commun. Math. Phys. {\bf 92}, 455 (1984).

\end{thebibliography}
\end{document}